\documentclass[reprint, superscriptaddress, amsmath,amssymb, aps, pra, prresearch, longbibliography, floatfix]{revtex4-2}

\usepackage{amsmath}
\usepackage{graphicx}
\usepackage{ulem}
\usepackage{soul}
\usepackage{xcolor}
\usepackage{bm}
\usepackage{dsfont}
\usepackage{tikz}
\usepackage{epsfig}
\usepackage{feynmf}
\usepackage{blindtext, rotating}
\usepackage{mathtools}
\usepackage{siunitx}
\usepackage{dsfont}
\usepackage{subcaption}
\usepackage{physics}
\usepackage{amsfonts}
\usepackage{xcolor}
\usepackage{ragged2e}
\usepackage{siunitx}
\usepackage{comment}
\usepackage{float} 
\usepackage{soul}
\usepackage{lipsum}
\usepackage{hyperref} 
\hypersetup{breaklinks=true, colorlinks=true, citecolor=blue, linkcolor=cyan, urlcolor=blue,filecolor=blue}

\usepackage{xcolor} 

\DeclareCaptionJustification{justified}{\justifying}

\DeclareSIUnit{\rad}{rad}

\begin{document}

\title{Levitated optomechanics with cylindrically polarized vortex beams}

\author{Felipe Almeida}
\email{felipe.almeida@ucl.ac.uk}
\affiliation{Department of Physics and Astronomy, University College London, Gower Street, WC1E 6BT London, UK}

\author{P. F. Barker}
\email{p.barker@ucl.ac.uk}
\affiliation{Department of Physics and Astronomy, University College London, Gower Street, WC1E 6BT London, UK}

\begin{abstract}
Optically levitated and cooled nanoparticles are a new quantum system whose application to the creation of non-classical states of motion and quantum sensing is fundamentally limited by recoil heating. Here we study the creation of stable 3D optical traps using optical cylindrically polarized vortex beams with radial and azimuthal polarization. We find that a significant reduction in recoil heating by up to an order of magnitude can be achieved when compared to conventional single Gaussian beam tweezers. We show that these beams allow trapping of larger particles outside the Rayleigh regime using both bright and dark tweezer trapping. By controlling the wavelength of the trapping laser, or the size of the particles, nonlinear and repulsive potentials of interest for the creation of non-classical states of motion can also be created.  

\end{abstract}
\maketitle

\section{Introduction}\label{sec:introduction}
The ability to control and cool the motion of objects while levitated in vacuum has led to the development of new large-mass quantum systems \cite{Deli2020}. Such systems are seen as promising candidates for exploring the mass limits of quantum mechanics\cite{RomeroIsart2010} and also for exploring the quantum nature of gravity \cite{bose2017}. While these systems are now seeing a growing number of applications in sensing\cite{gosling2024,wang2024}, significant research efforts are focusing on the creation of non-classical states of motion, including spatial superpositions, squeezed mechanical states and Fock states \cite{Weiss2021,Bateman2014, ernotk2020, Arndt2014, Rashid2017}.  

While the manipulation and cooling of these levitated systems is well developed, their use is currently limited for some applications by motional decoherence\cite{Romero-Isart2011} induced by the recoil of scattered photons \cite{Jain2016}. This places significant constraints on the timescales over which non-classical states can be created, manipulated and witnessed \cite{Magrini2021, Tebbenjohanns2021}. As such, protocols to create these states, for example by rapid wavefunction expansion, must be carried out over short timescales, well below the decoherence time, or in the absence of light when the particle will free fall. While electrical and magnetic levitation do not suffer from this type of decoherence, the higher trap frequencies associated with optical trapping, coupled with the ability to rapidly manipulate the trapping potential, make optical levitation one of the most attractive platforms for the creation of massive quantum systems. 

Alternative optical schemes to reduce decoherence from photon recoil can also in principle be explored via the engineering of optical potentials\cite{Dago2024}. For example, the use of non-classical light sources for trapping and readout \cite{GonzalezBallestero2023} and the use of a reflective boundary \cite{Gajewski2025} have been studied. 

Most optical trapping experiments use tightly focused Gaussian beams, typically with linear polarization. These conveniently create harmonic potentials near the trap center with different frequencies in each direction. 
Other beam geometries, often referred to as structured light \cite{Forbes2021, Yang2021}, use more complex fields such as higher order modes to create nonlinear trapping potentials\cite{Arita2017, Almeida2023, Dago2024, Tandeitnik2024}, or to induce orbital angular momentum \cite{Allen1992, GarcsChvez2003, Simpson1997, ONeil2002}. The ability to modulate the polarization in the transverse plane enables a more general class of light fields, known as vector beams \cite{Yang2021}. Those with cylindrical symmetry in polarization \cite{Khonina2019, Maurer2007, Zhan2009} can be used to create more tightly focused beams when compared to Gaussian beams. They can also be used to create "donut" beams with a central intensity minimum, both of which can in principle be used to decrease recoil heating of particles trapped within their focus.

In this paper, we study the use of these cylindrically polarized vortex beams as tools to reduce photon recoil and bulk heating, and for engineering both confining and inverting potentials. We study the effect of Mie resonances on optical forces for a high refractive index material, using silicon as an example ($\mathrm{Si}$) and compare this with conventional trapping of SiO$_2$ in a Gaussian beam. We show that three-dimensional trapping is available for both radial and azimuthal polarized fields constructed from a single focused beam. Our calculations also indicate that the photon recoil heating from some beam configurations can be significantly smaller than that of a Gaussian beam, even in the Rayleigh regime, by up to an order of magnitude. 
\section{Trapping in Gaussian beams}
Before discussing trapping in vector vortex beams, we describe trapping in Gaussian beams (GB) to illustrate the effects of Mie resonances in high refractive index nanoparticles.
The variation of trapping forces can be conveniently described in terms of the size parameter defined to be, $kR=2\pi R/\lambda$, where $R$ and $\lambda$ are the radius of the particle to be trapped and wavelength of the optical field, respectively. Optical forces can be roughly divided into the Rayleigh regime where $kR |n-1| \ll 1$ and the Mie scattering regime where $kR |n-1| \approx 1$, $n$ being the refractive index of the particle. 

We calculate optical forces using the optical tweezers toolbox (OTT)\cite{Nieminen2007} using the incident and the resulting scattering fields from the particle. OTT uses the T-matrix formalism within a vector spherical wave basis, providing a vectorial and non-paraxial treatment of both the incident and scattered electromagnetic fields. 
This makes it particularly well-suited for Mie scattering 
where the analytical solutions are not available. 
The forces and optical potentials are calculated using the parameters shown in Table \ref{tab:parameters}. This includes the power, wavelength, particle refractive index, particle mass density and numerical aperture (NA).  

\begin{figure}[h]
    \centering
    \includegraphics{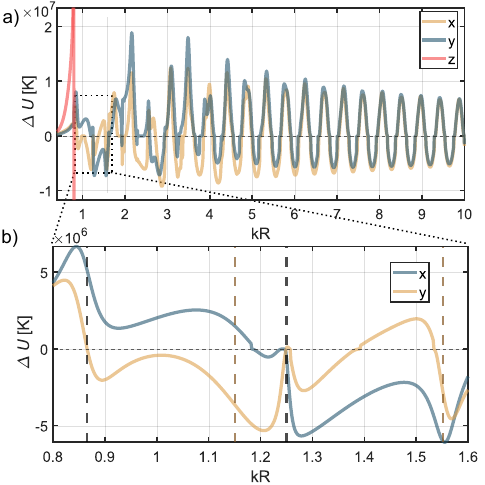}
    \caption{a) Plots of trap depth, $\Delta U$, along the three trap axes ($x, y, z$) as a function of size parameter $kR$ for a Si spherical nanoparticle in a linear polarized Gaussian beam (GB). The depth oscillates between attractive (positive) and repulsive (negative) in the Mie Regime for a size parameter $kR \approx 1$. b) Plots of trap depth over a reduced size parameter interval shows asymmetry between the $x$ and $y$ potential in the transition from the Rayleigh to Mie Regime. Black and brown dashed lines represent the location of TE and TM modes respectively.}
    \label{fig:3d_bright_trapping}
\end{figure}

The optical trap depth $\Delta U$ for a linearly polarized GB trap, propagating along the positive $z$ axis, is shown for each direction ($x,y,z$) in Fig \ref{fig:3d_bright_trapping}. For each particle radius, $R$, we first record the stable equilibrium point along the axial direction, $z_{\mathrm{eq}}$, corresponding to the potential minimum. We then determine the local potential maxima on either side of this equilibrium point. The lower of these two maxima, located at $z=z_{\mathrm{max}}$, determines the depth of the axial trap given by $\Delta U_z = U(0,0,z_{\text{max}}) -  U (0,0, z_{\text{eq}}) >0$. The transverse trap depth is calculated at $\bm{r}=(0,0,z_{\text{eq}})$ by $\Delta U_{x} = U(x_{max},0,z_{\text{eq}}) -  U (0,0, z_{\text{eq}}) >0$  where $x_{max}$ is the position where the potential is maximum. If no equilibrium point is found in $z$ direction,  we calculate the axial and transverse potential at the beam focus ($\mathrm{z_{\text{eq}}}=0$) and set the axial trap depth to be negative which occasionally results in discontinuities in calculated potential well depth $\Delta U$. Our calculations for the GB are in good agreement with the analytical values in the Rayleigh regime. The location of the trapping peaks is consistent with other Mie trapping calculations \cite{Stilgoe2008}.
\begin{table}[h!]
\centering
\caption{Table of parameters used in the simulations of optical forces on silicon and silica nanoparticles. }
\label{tab:parameters}
\begin{tabular}{l l}
\hline
\textbf{Parameters} & \textbf{Description} \\
\hline
$\lambda =\SI{1550}{\nano\meter}$ & Laser wavelength \\
${\mathrm{NA}} = 0.8$& Numerical aperture \\
${\mathrm{P}} = \SI{500}{\mW}$& Laser power \\
$n_{\mathrm{Si}} = 3.48  +  i \; 5.3\times 10^{-11} $ & Refractive index of $\mathrm{Si}$ at $\lambda$  \\
$n_{\mathrm{SiO_2}} = 1.46 + i \; 5\times 10^{-9} $ & Refractive index of $\mathrm{SiO_2}$ at $\lambda$  \\
$\rho_{\mathrm{Si}} = \SI{2200}{\kg \per \meter^{3} }$ & Mass density of $\mathrm{Si}$ \\
$\rho_{\mathrm{SiO_2}}$ = \SI{1850}{\kg \per \meter^{3}}& Mass density of $\mathrm{SiO_2}$ \\
\hline
\end{tabular}
\end{table}

Figure \ref{fig:3d_bright_trapping}a shows that the trap depth oscillates between positive (attractive) and repulsive (negative) values with increasing size parameter. Repulsive trapping in any axis implies that a particle cannot be stably trapped in 3D. Three-dimensional trapping of Si becomes impossible in a GB as the force in the $z$ direction becomes dramatically large in the Mie Regime for $kR > 0.8$ as either the scattering force dominates over the gradient force in the $z$ direction or gradient force becomes repulsive in the transverse direction. 

Figure \ref{fig:3d_bright_trapping}b is a zoomed-in plot highlighting the variation in the transverse potential near $kR\approx 1$ at the focal plane illustrating how the transverse potential can be repulsive or attractive (trapping). Also shown are the locations of the first transverse electric (TE) and transverse magnetic (TM) modes shown as the black and brown dashed lines respectively.  These modes are central to both increased scattering forces in the $z$ direction and to the repulsive or attractive dipole forces in the transverse directions $(x,y)$. The location of these resonances is determined through Mie theory and is dependent on the size parameter, while their excitation is dependent on the polarization of incident light \cite{Zhang2023, Fontes2005, Ng2008, Kislov2021}.

The origin of the oscillations in the sign of the optical forces 
is well known and has been studied extensively for a range of fields \cite{Kislov2021, Chen2011} from focused Gaussian beams \cite{Stilgoe2008} to evanescent fields \cite{Li2025}. The phase of the field excited by the Mie/WGM resonance changes with size parameter and goes from being in phase below the resonance to out of phase above it, leading to a change in the sign of the gradient force \cite{Li2025}. 
Multiple nearby adjacent resonances can lead to a complex landscape of repulsive and attractive forces and this is observed as the size parameter increases. Both TM and TE resonances are excited by a linearly polarized Gaussian beam as the size parameter increases. This leads to an attractive gradient force in one transverse direction with repulsion in another, as shown in Fig. \ref{fig:3d_bright_trapping}b around the first few TE and TM resonances. This behavior suggests that the use of beams with cylindrical polarization symmetry, such as cylindrical vector vortex beams, offers a route to stable trapping in the transverse plane since the same Mie resonance could be excited in both $x$ and $y$ directions simultaneously.  For example, for an azimuthally polarized beam, predominant excitation of TE modes would occur while for radially polarized beams TM excitation would predominate \cite{Woniak2015}. Additionally, in the case of azimuthally polarized beams which are characterized by a dark spot in the center of the beam, a reduction in both bulk heating and recoil heating may be feasible.    

\section{Cylindrically polarized vector vortex beams} \label{sec:CP_beam}

\begin{figure}[h]
    \centering
\includegraphics{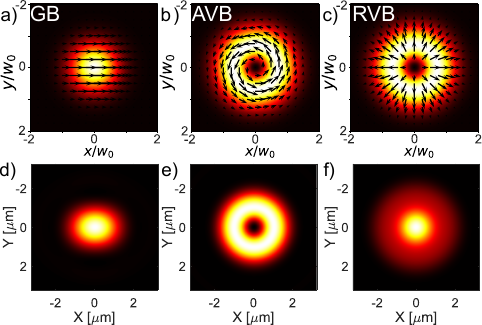}
    \caption{ Plots of the spatial intensity and polarization profile for a) Gaussian beam (GB), b) azimuthal vector beam (AVB) and c) radial vector beam (RVB). d-f) are the intensity profiles of these beams when focused by a $\text{NA}=0.8$ lens.}
    \label{fig:beam_profile}
\end{figure}
The vector vortex beams used as inputs for calculating the optical forces in OTT are represented in the paraxial limit as a superposition of Hermite-Gauss (HG) optical modes. The corresponding azimuthal vector electric field is given by:
\begin{equation}
\label{eq:eq_azi}    
    \begin{aligned}
        \text{E}_{\mathrm{azi}} =\sqrt{\frac{P}{\epsilon_0c}} \left( \frac{u^{HG}_{10} \bm{\hat{y}} - u^{HG}_{01}\bm{\hat{x}}}{\sqrt{2}}  \right )= {\text{E}_{00}^{HG}}  e^{-i \xi(z)} \begin{bmatrix} y  \\ -x  \end{bmatrix} 
    \end{aligned}    
\end{equation}
\noindent and the radial vector electric field is:
\begin{equation}
\label{eq:eq_rvb}    
    \begin{aligned}
        \text{E}_{\mathrm{rad}} = \sqrt{\frac{P}{\epsilon_0c}} \left ( \frac{u^{HG}_{10} \bm{\hat{x}} + u^{HG}_{01}\bm{\hat{y}}}{\sqrt{2}} \right )= {\text{E}_{00}^{HG}}  e^{-i \xi(z)} \begin{bmatrix} x  \\ y   \end{bmatrix} ,
    \end{aligned}    
\end{equation}

\noindent where $u_{mn}^{HG}$ are the normalized Hermite-Gauss modes for orders $m$ and $n$ \cite{daSilva2020}, ${\text{E}_{00}^{HG}}$ is the electric field of a Gaussian beam for $m=n=0$ and where $w(z)$ and $\xi(z)$ are the beam waist and Gouy phase, respectively, defined as:
\vspace{-4pt}
\begin{subequations}
\begin{gather}
    \text{E}_{00}^{HG} = \sqrt{\frac{P}{\epsilon_0 c}} \frac{1}{ \sqrt{2\pi} \; w(z)} \exp \left (\frac{\rho^2}
    {w(z)^2}  \right )  \exp \left(\frac{-ik \rho^2}{2R}  - i\xi(z)\right )  \\  
w(z) =  w_0\sqrt{1+ \left(\frac{z}{z_R}\right)^2} \\
\xi(z) =  \arctan \left (\frac{z}{z_R} \right)
\end{gather}    
\end{subequations}
\noindent where $w_0=\lambda/\pi \mathrm{(NA)}$ and $z_R= n_m\lambda/\pi \mathrm{(NA)}^2$ are the beam waist and Rayleigh range, respectively. The wavenumber is given by $k=2\pi n_m/\lambda$, and the wavefront radius by $R(z)=z(1+z_R^2/z^2)$. The resulting transverse intensity profiles of the GB, the AVB and the RVB can be seen in the first row of Fig. \ref{fig:beam_profile}a-c plotted with the respective polarization vectors.

To calculate the optical fields within the focal region produced by a high numerical aperture (NA) lens, we employ a vector spherical wave function expansion. The corresponding transverse intensity profiles of the focused GB, AVB and RVB are shown in Fig. 2d-f).
Due to the non-paraxial nature of the focusing, the transverse polarization components are projected onto the propagation/axial direction.   For the RVB, this creates a strong electric field component in the axial direction directly changing its intensity profile to a bright focal point, While for the AVB the central intensity minimum is maintained at the focus. The relative intensity profiles for the focused GB, AVB and RVB are shown in Fig. \ref{fig:beam_profile}d-f). 
It is well known that a RVB can be focused to a spot size significantly smaller than a GB \cite{Dorn2003}, creating an optical potential that is narrower (Fig. \ref{fig:beam_profile}f). This has been verified through the enhanced trap stiffness in optical tweezers experiments \cite{Kozawa2010, Huang2012} and more recently via optical levitation experiments \cite{Michihata2009}, when compared to GB traps. 
In contrast, an AVB generates a purely transverse electric field with a null electric field component in the axial direction at the center of the focused beam. This is responsible for maintaining an annular (donut) intensity profile, even at a high numerical aperture \cite{Zeng2017}, as seen in Fig. \ref{fig:beam_profile}e. This beam has also been used for optical trapping within the annular region \cite{Guo2024}. 

In the next two sections, we 
discuss stable trapping scenarios for high ($\mathrm{Si}$) and low ($\mathrm{SiO_2}$) refractive index nanoparticles in the cylindrically polarized vector vortex beams. We begin in section IV by considering optical trapping in the dark center of an AVB (dark trapping), as this configuration illustrates the role of excited Mie resonances in high refractive index particles. The discussion is divided into two subsections, where we explore two different configurations with i) two counter propagating, low numerical aperture beams and ii) a highly focused single-beam. In section V, we turn to trapping of Si in the RVB, where particles are confined in the region of maximum intensity of light (bright trapping). Unlike the AVB case, the RVB enables stable trapping even in the Rayleigh regime. 
The tight focusing achievable with a RVB can produce steeper optical potentials and consequently higher oscillation frequencies when compared to a conventional GB. 
\section{Trapping in azimuthallly polarized vortex beams} \label{sec:trapping_in_AVB}
\subsection{Low numerical aperture trapping in counter propagating beams}
\begin{figure}[h]
    \centering
    \includegraphics{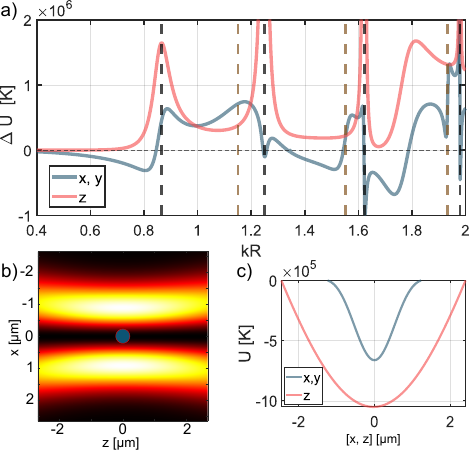}
    \caption{a) Calculated trap depths, $\Delta U$, as a function of size parameter for a  counter propagating AVB trap using focusing lenses of numerical aperture NA= 0.4. The axial force becomes attractive for the entire range, with maxima at the TE modes. Black and brown dashed lines represent TE and TM modes, respectively. b) The relative intensity profile of the counter propagating AVB beam trap. The location of the trapped nanoparticle is represented by the grey circle of size $R=\SI{222} {nm}$. c) The optical potential in the transverse ($x,y$) and longitudinal directions ($z$) for size parameter $kR=0.9$}
    \label{fig:3d_CP_trapping}
\end{figure}

To illustrate the trapping mechanism in the cylindrically polarized vector vortex beams we start by studying the role of resonances for the simple case of low numerical aperture beams (NA = 0.4) trapping a silicon nanosphere. Although previous experiments have shown optical trapping of Rayleigh particles in the annular ring of an AVB \cite{Guo2024}, we are interested in Mie particles which can be trapped in the dark center of the AVB, where we expect reduced recoil heating \cite{Almeida2023}. Therefore, we perform trap depth calculations within this region of interest, defined as the interval between the two bright lobes of the annular beam $[\frac{-w_0}{\sqrt2} ,\frac{w_0}{\sqrt2}]$. 
As in the case for low numerical aperture GBs the scattering force is always larger than the gradient force along the $z$ axis and therefore no stable equilibrium exists for a single beam. We therefore consider the case of two counter propagating beams along the $z$ axis such that there is always stable trapping in this direction. These two counter propagating beams are separated in frequency by a few MHz so that any interference between them does not play a role in the trapping along the $z$ direction.  We calculate the optical potential as a function of the size parameter with the same protocol described in section II and the results are shown in Figure \ref{fig:3d_CP_trapping}a. Note that the $y$ axis is the same as the $x$ axis plot due to symmetry and is therefore not shown here. As expected, the optical potential is always positive along the $z$ direction, with a maximum at each of TE resonances shown as the black vertical dotted lines. The transverse potential along the $x$ axis exhibits the expected gradient force behavior near a resonance. It is repulsive (non-trapping) on the left side and attractive (trapping) on the right side of the first TE resonance at $kR=0.867$.
While the trapping is dominated by the TE resonances, TM resonances are also excited at larger size parameters. This occurs because as the particle moves away from its transverse equilibrium the azimuthal polarization has a small radial component with respect to the surface of the sphere and TM modes can be excited. 
Figure \ref{fig:3d_CP_trapping}b illustrates the equilibrium position of a $kR=0.9$ $(R=\SI{222} {nm})$ particle within the resulting optical field. Fig. \ref{fig:3d_CP_trapping}c shows the optical potential created by the counter propagating AVBs.

\subsection{High numerical aperture trapping in a single-beam}
\begin{figure}[h]
    \centering
    \includegraphics{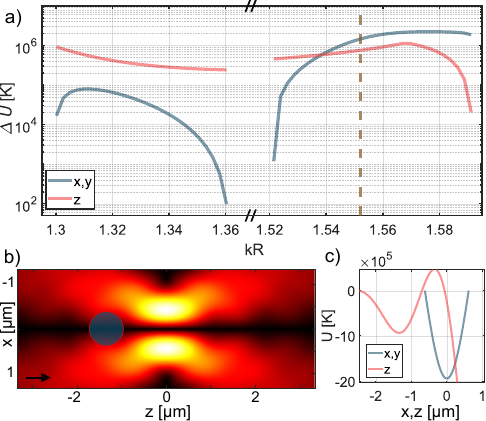}
    \caption{a) Calculated trap depth, $\Delta U$, for the single beam AVB focused using a 0.8 NA lens. The axial repulsive force becomes dominant near the TE modes. Three-dimensional trapping is found for size parameter regions given by $kR=[1.30, 1.36]$ and $[1.52,1.59]$. 
    Black and brown dashed lines represent TE and TM modes, respectively. b) Relative intensity profile of AVB single beam trap. The circle represents the trapped particle of radius $R=\SI{385}{nm}$ at the equilibrium location $z=z_{\text{eq}} =\SI{-1.3}{\micro m}$. c) The optical potential for $kR=1.56$ ($R=\SI{385}{nm}$) at the equilibrium location.}
    \label{fig:3d_dark_trapping}
\end{figure}

We now explore trapping with a single AVB created by focusing with a high NA lens. Here again we consider the trapping of silicon nanospheres and the beam propagates along the positive $z$ axis. The trapping parameters are given in Table \ref{tab:parameters}. Figure \ref{fig:3d_dark_trapping}a shows the calculated trap depths in all directions for a single AVB as a function of the size parameter. This shows the regions where 3D trapping is possible and is over two intervals, $kR=[1.30, 1.36],[1.52,1.59]$. At the limit of these two intervals either the axial or transverse trap depth goes to zero even though the finite resolution of our calculations does not resolve this.
Figure \ref{fig:3d_dark_trapping}b shows the particle equilibrium within the AVB found for a $kR=1.56$ ($R=\SI{385}{nm}$) particle. The particle finds an equilibrium before the focal point, indicating the presence of an optical pulling force \cite{Chen2011}. 
The optical potentials can be seen in Figure \ref{fig:3d_dark_trapping}c. The variations in the sign of the $x$ axis trap depth are once again explained in terms of the interaction of high order modes \cite{Zhang2023}, where the size parameter increases and passes through the resonance. The phase change across resonance leads to a change in the sign of the induced force or the effective polarizability\cite{ Chen2011, Lepeshov2023}. 

\begin{figure}
    \centering
    \includegraphics{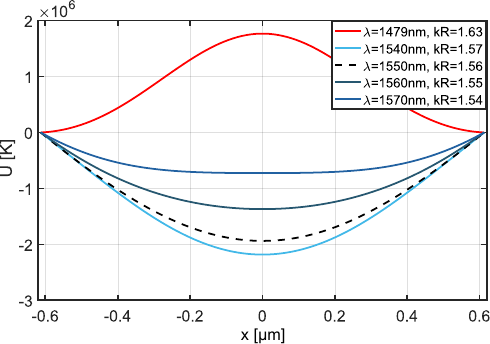}
    \caption{Plots of the transverse potential profile as a function of wavelength in an AVB trap. Curves are shown for a particle radius $R=\SI{385}{\nano\meter}$. Blue shifting the trapping wavelength flattens the potential and allows the creation of repulsive potentials.}
    \label{fig:potential_for_wavelength}
\end{figure}

The single beam AVB allows us to tune the transverse potential in this direction by changing the size parameter through the wavelength of the optical field. Figure \ref{fig:potential_for_wavelength} shows the transverse potential calculated for Si nanoparticles with $R=\SI{385}{\nano\meter}$ at different wavelengths. The black dotted lines are the calculated potential for $\lambda=\SI{1550}{\nano\meter}$. Blue shifting the trapping wavelength flattens the potential indicated by the curves at $\lambda=\SI{1560}{\nano\meter}$ and $\lambda=\SI{1570}{\nano\meter}$. These plots indicate that changing wavelength allows tuning of the potential from a harmonic to nonlinear and even to a repulsive potential ($\lambda=\SI{1479}{\nano\meter}$). Inverted and nonlinear potentials \cite{Tomassi2025,RieraCampeny2024} can be used for fast wavefunction expansion and have been proposed for evidencing non-classical motion. Such potentials generally require fast switching which could be accomplished through established electro-optic control \cite{Bonvin2024, Rashid2016}. This method could also be used to switch between AVB and other beams on timescales much faster ($<$ 1 ns) than the particle's trap dynamics on the $\mu s$ timescale. 

\section{Trapping in radially polarized beams}

It is well known that vortex beams with radial polarization can be more tightly focused than a GB for high NA \cite{Dorn2003}. As a radially focused beam has a larger intensity gradient in both the transverse and axial directions, less intensity is required to create the same trapping frequencies when compared with a GB trap.  This suggests that this type of trap may have lower recoil, since a lower optical field intensity can be employed to produce a trap of the same frequencies while maintaining the bulk heating at least at the value of a GB trap.

\begin{figure}[h]
    \centering
    \includegraphics{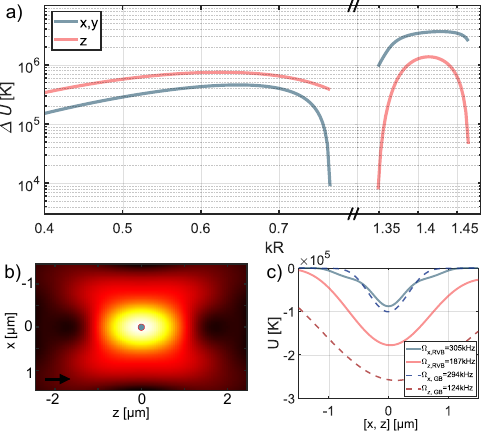}
    \caption{a) Plots of trap depth calculated for $\mathrm{Si}$ nanoparticle in a single beam RVB as a function of size parameter. The axial repulsive force becomes dominant near the TM modes. Three-dimensional trapping occurs between $kR=[0.4, 0.76], [1.35, 1.46]$.  The black and brown dashed lines represent the TE and TM modes, respectively. b) A plot of the relative intensity profile of the RVB trap. A circle with radius $\SI{74}{nm}$ ($kR=0.3$) is drawn at the equilibrium trapping position. c) The optical potential for $P=\SI{500}{mW}$ for $R=\SI{74}{nm}$. The RVB potential is smaller than that of the GB. The trap frequencies for the GB ($\Omega_{G,x}$ $\Omega_{G,z}$) and the RVB ($\Omega_{R,x}$, $\Omega_{R,z}$) are shown for the $x$ and $z$ trap axes.  }
    \label{fig:3d_RVB_trapping}
\end{figure}

Figure \ref{fig:3d_RVB_trapping}a shows plots of the calculated trap depth for spherical $\mathrm{Si}$ nanoparticles levitated within an RVB, propagating in the positive $z$ direction, with parameters from Table \ref{tab:parameters}. The curves show three-dimensional trapping at the Rayleigh limit $kR < 0.76$ and a second interval within Mie Regime at $kR=[1.35, 1.46]$. 

Figure \ref{fig:3d_RVB_trapping}b illustrates the equilibrium location within the intensity profile of the focused beam for a $\mathrm{Si}$ nanoparticle where  $kR=0.3$ $(R=\SI{74}{\nm})$. Figure \ref{fig:3d_RVB_trapping}c shows the corresponding axial and transverse optical potentials for GB and RVB traps with their corresponding mechanical frequencies when the same power is used. The transverse potential is tighter than that of the GB near the focal point due to the intensity distribution of the RVB at high NA. As expected, the mechanical frequencies are higher for the RVB than for the GB trap, and in particular for the axial direction due to the tighter focus of the RVB.

\section{Recoil heating}

Recoil or shot noise heating results from the scattering of photons by the trapped particle and the subsequent diffusion/heating of its trapped motion. This occurs for both translational and rotational motion but here we focus on the translation heating.

For a spherical Rayleigh particle the translational shot noise can be calculated from the average change in energy due to the scattering of an incoming photon $\bm{\hat{k}_i}$. 
The probability of an incoming photon to be scattered into a solid angle $d\Omega$  is given by \cite{Seberson2020} 
$    P(\bm{\hat{k}_s})d\Omega = \frac{1}{\sigma} \left( \frac{d\sigma}{d\Omega} \right) d\Omega $
 where $\bm{\hat{k}_s}$ is the scattered direction, $d\sigma/d\Omega$ is the differential scattering cross section and the probability distribution is normalized such that $\int P(\bm{\hat{k}_s})d\Omega=1$. The average change in energy $\langle\Delta E_j\rangle$ due to a single photon is then
\begin{equation}
    \langle\Delta E_j\rangle =  \epsilon\int_{\Omega} P(\bm{\hat{k}_s}) \left( {k}_{i,j} - {k}_{s,j} \right )^2  d\Omega,
\end{equation}
\noindent where an incoming wave traveling along the optical axis $\bm{\hat{k}_{i}}=\bm{\hat{z}}$ is assumed and the scattered wave direction $\bm{\hat{k}_{s}}$, whose Cartesian components, marked by subscript $j={x,y,z}$, ${{k}_{s,j}}$ are written in spherical coordinates: $k_{s,x}=\sin \theta \cos \phi$, $k_{s,y}=\sin \theta \sin \phi$, $k_{s,z}=\cos \theta $ and $\epsilon=\frac{\hbar^2k^2}{2m}$ is the recoil energy associated with the scattering of a single photon.  The total translational energy transfer is then $\dot{E}_R = \frac{I_0 \sigma_{\text{scat}}}  { \hbar \omega_0 }\langle  \Delta E\rangle_R $  where $I_0$, $\sigma_{\text{scat}}$ and $\omega_0$ are the laser intensity, scattering cross section and laser frequency, respectively, and $\langle  \Delta E\rangle_R  = \sum_j \Delta E_j =2\epsilon$.

As we reach the domain of Mie scattering theory \cite{Maurer2023}, we follow Ref. \cite{Seberson2020} where the total translational energy gained per second due to scattering is given by
\begin{equation}
    \dot E_{j} = \frac{\epsilon_0 c \hbar }{4m \omega} \int \left | \frac{\partial \bm{\mathcal{E}_{k, k'}} (\theta, \phi)}{\partial j} \right |_{x_i=0} ^2  \left( {{k}_{i,j}} - {{k}_{s,j}} \right )^2  d\Omega,
    \label{eq:Mie_heating}
\end{equation}
\noindent where $\bm{\mathcal{E}_{ k,k'}} (\theta, \phi) = {\mathcal{E}_{\theta}}  (\theta, \phi) \bm{\hat{\theta}} + {\mathcal{E}_{\phi}}  (\theta, \phi) \bm{\hat{\phi}}$ is 
the far field scattering amplitude into a given direction $\bm{\hat{k}_{s}}$. The numerical derivatives were evaluated from the scattered fields calculated with the OTT as discussed in appendix A. These are consistent with the GB in Ref. \cite{Seberson2020}. 
We define the photon recoil heating rate in direction $j$ as the ratio between
the mechanical energy gained per unit time over the energy of a single mechanical
quanta as: 
\begin{equation}
\label{eq:recoil_heating}
    \Gamma_j = \frac{I_0 \sigma_{\text{scat}}}{\hbar \omega_0} \frac{\langle \Delta E_j \rangle }{\hbar \Omega_j} = \frac{\dot{E}_{j}}{\hbar \Omega_j}  ,
\end{equation}
\noindent where $\Omega_j$ is the oscillation frequency of the levitated nanoparticle in the $j$ direction.  
In the remaining part of this section we 
consider the photon recoil heating that results from each of the AVB and RVB trap geometries for Si. We compare it to the trapping of $\mathrm{SiO_2}$ nanoparticle in a GB with the same power, NA for single or counter propagating beams.
\subsection{Recoil heating in azimuthal vector beams}\label{sec:recoil_AVB}
\begin{figure*}[ht!]
    \centering
    \includegraphics{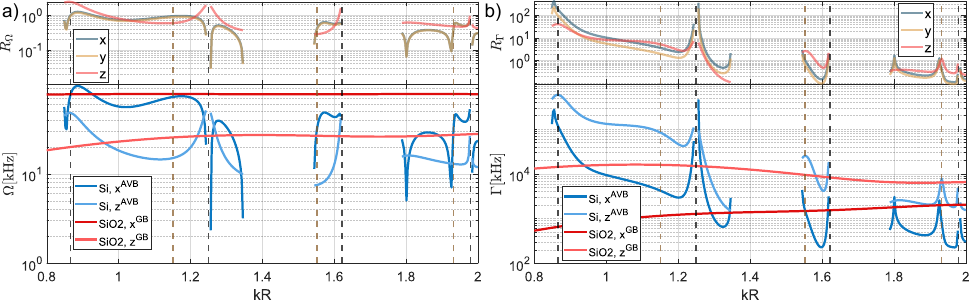}
    \caption{Comparison of the trapping of Si and $\mathrm{SiO_2}$ in counter propagating AVB and GB, respectively.  a) Trap oscillation frequencies (bottom), with the corresponding frequency ratio, $R_\Omega = \Omega_{\text{AVB}}/\Omega_{\text{GB}}$ shown for x,y,z directions (top). b) Recoil heating rate (bottom) and the correspnding ratio, $R_\Gamma = \Gamma_{\text{AVB}}/\Gamma_{\text{GB}}$, with respect to the counter propagating GB trap. The curves in a) indicate weaker trapping in $x$ and $y$ directions from the AVB trap when compared with the GB trap and a maximum reduction in recoil heating of $10.6$ at $kR=1.97$}
    \label{fig:recoil_heating_AVB_cp}
\end{figure*}

\subsubsection{Low numerical aperture trapping in counter propagating beams}
$\mathrm{SiO_2}$ has been widely used for optical levitation using GBs. It has mechanical and chemical stability under high vacuum and a low absorption coefficient near IR trapping wavelengths. Its relatively low refractive index means that in the Rayleigh regime it scatters less power than higher refractive index materials such as silicon and therefore suffers less recoil heating. However, as seen in section \ref{sec:trapping_in_AVB}, a high refractive index material enables trapping within the dark spot of an optical beam, leading to an overall reduction in photon recoil heating. 

Figure  \ref{fig:recoil_heating_AVB_cp} shows the calculated trap oscillation frequencies and photon recoil heating in the $x$ and $z$ directions as a function of the size parameter for trapped Si nanoparticles. We compare these values to those calculated for trapping of $\mathrm{SiO_2}$ nanoparticles in counter propagating GB trap with the same NA. 
Also shown are the ratios of these frequencies from each trap. The plots for Si in AVB show a strong dependency of the oscillation frequencies around the TE modes (black vertical dotted lines).  Figure  \ref{fig:recoil_heating_AVB_cp}b shows the recoil heating rates and the ratio with respect to the GB trap. We find that the recoil heating ratio reduces when compared to the counter propagating GB trap as we increase particle size with a reduction by factor of $10.2$ when $kR=1.6$ and $10.6$ at $kR=1.97$. 
\subsubsection{High numerical aperture trapping in a single-beam}
\begin{figure}[h]
    \centering
    \includegraphics{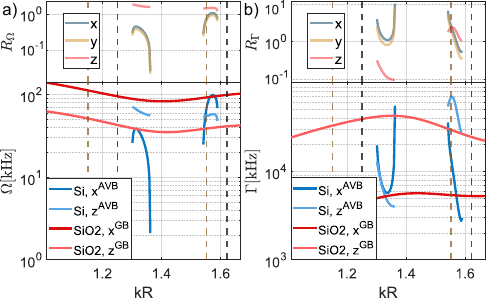}
    \caption {Comparison of Si and $\mathrm{SiO_2}$ in AVB and GB traps, respectively.  a) Trap oscillation frequencies (bottom), with the corresponding frequency ratio, $R_\Omega = \Omega_{\text{AVB}}/\Omega_{\text{GB}}$ shown for x,y,z directions (top). b) Recoil heating rate (bottom) and the corresponding ratio,$R_\Gamma = \Gamma_{\text{AVB}}/\Gamma_{\text{GB}}$, with respect to the GB trap. Stable trapping points are found within the dark focus for $kR=[1.30, 1.36],[1.52,1.59] $. The plots show a weaker trapping in $x$ and $y$ directions and a maximum reduction in recoil heating of $6.25$ at $kR=1.36$ relative to GB. }
    \label{fig:recoil_heating_azimuthal}
\end{figure}

We now discuss trapping for high refractive index $\mathrm{Si}$ in the single AVB.  We compare this to the trapping of $\mathrm{SiO_2}$ particles using a single GB. We do not compare with the trapping of Si in GB because as shown in Fig. \ref{fig:3d_bright_trapping} the Mie resonances preclude the trapping of Si by a GB in this size parameter regime.
We calculated oscillation frequencies, photon recoil heating rates and their respective ratios over the size parameter range $kR=[0.4, 2]$. Figure \ref{fig:recoil_heating_azimuthal}a shows these values over a smaller range in which three-dimensional trapping is found corresponding to the intervals $kR=[1.30,1.36],[1.52,1.59]$. Note that the AVB trapping frequencies in the $x$ and $y$ directions are the same and therefore we only show $x$. 
Generally, we observe two different behaviors for these trapping intervals. In the first trapping interval the transversal trapping frequencies in the AVB are lower than for the GB, but the axial frequencies are approximately two times larger. In the second trapping interval, the transverse and axial trapping frequencies are comparable and also more constant. The recoil heating rate and ratios for these two beams are shown in Fig. \ref{fig:recoil_heating_azimuthal}b again presenting two different behaviors. The first interval presents a considerable reduction by up to a factor of $6.25$ in the recoil heating rate at $kR \approx 1.36$ in the axial direction at the cost of a slightly higher recoil heating in the transverse directions.
In the second interval, heating in the axial direction is always larger than the GB but in the transverse directions the heating rates transition from being higher to lower across the TM resonance.

\subsection{Recoil heating in radial vector beams}

\begin{figure}[h]
    \centering
    \includegraphics[width=8.6cm]{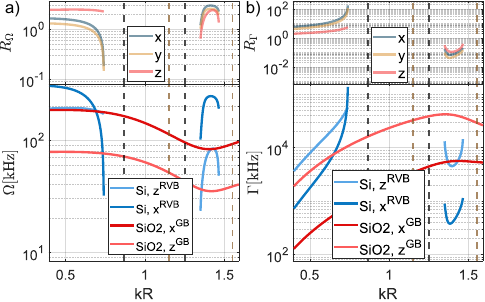}
    \caption{ Comparison of Si and $\mathrm{SiO_2}$ in RVB and GB traps, respectively.  a) Trap oscillation frequencies (bottom), with the corresponding frequency ratio, $R_\Omega = \Omega_{\text{RVB}}/\Omega_{\text{GB}}$ shown for x,y,z directions (top). b) Recoil heating rate (bottom) and the corresponding ratio, $R_\Gamma = \Gamma_{\text{AVB}}/\Gamma_{\text{GB}}$, with respect to the GB trap. Stable trapping points are found for $kR=[0.4,0.76],[1.35, 1.46]$ for the RVB. The second interval shows less recoil heating than that of a $\mathrm{SiO_2}$ nanoparticle in a GB with recoil heating reduced by a factor of $9.2$ at $kR=1.39$.}
\label{fig:recoil_heating_radial_Si_SiO2}
\end{figure}

We now discuss trapping for high refractive index Si in the RVB. 
We calculated oscillation frequencies and photon recoil heating rates over the same size parameter range as in section \ref{sec:recoil_AVB}. 
As before, we compare these results with those for $\mathrm{SiO_2}$ in GB with the corresponding ratios shown in the upper panels.
Figures \ref{fig:recoil_heating_radial_Si_SiO2}a and \ref{fig:recoil_heating_radial_Si_SiO2}b show those values for both trap geometries. We show the reduced interval $kR=[0.4,0.76],[1.35, 1.46]$ where three-dimensional trapping for $\mathrm{Si}$ is found for RVB. Note that although the first interval (Rayleigh regime) presents a higher oscillation frequency in all three directions for the RVB within $kR<0.65$, the recoil heating ratio is higher for RVB than GB due to its larger scattering cross section $\sigma_{\text{scat}}$. 
A second interval appears between the two Mie resonances corresponding to the TM and TE resonances. In this interval, recoil heating in the axial direction is reduced by up to a factor of $9.2$ at $kR=1.39$ also increasing the oscillation frequency ratio by a factor of $2$.



\section{Bulk temperature}
Stable optical levitation requires an internal temperature that is less than the melting point of the particle.
It is therefore important to estimate the internal temperature of the nanosphere in the different optical trapping configurations considered in this work. We determine the equilibrium bulk temperature $T$ at which the power absorbed by the particle (laser and black body radiation) is balanced by the power emitted by the particle (black body emission and gas collisions) such that
\begin{equation}
 P_{\mathrm{abs}} + P_{\mathrm{BB, abs}} - P_{\mathrm{BB, emit}}(T) - P_{\mathrm{gas}} (T)=0. 
 \label{eq:T_min}
\end{equation}

Here $P_{\text{abs}}$ is the absorbed laser power, $P_{\mathrm{BB, abs}}$ is the power absorbed from the environment, $P_{\mathrm{BB, emit}}$ is the black body power emitted by the object and  $P_{\mathrm{gas}}$ denotes the heat removed from the nanoparticle by the residual background gas. The equilibrium temperature is found from Eq. \ref{eq:T_min} by using numerical root finding.

The absorbed power is critically dependent on the complex refractive index, or absorption coefficient, and initially we model it as constant, independent of laser power. Following Ref. \cite{Lepeshov2023}, we define $P_{\text{abs}} = P_{\text{inc}}-P_{\text{out}}$ for incident power $P_{\text{inc}} = \frac{1}{2} \int_{S} ( \bm{E_{inc}}  \times  \bm{H_{inc}}^*  )\cdot d\bm{S} $ normalized to input power (Table \ref{tab:parameters}) and the outgoing power $P_{\text{out}} = \frac{1}{2} \int_{S} (\bm{E_{inc}} + \bm{E_{scat}}) \times  ( \bm{H_{inc}}^* + \bm{H_{scat}}^* ) \cdot d\bm{S} $ where $\bm{S}$ represents a sphere enclosing the nanoparticle, that is positioned at the focal plane. The black body emitted power, $P_{\mathrm{BB, emit}}$, is given by 
\begin{equation}
        P_{\mathrm{BB, emit}}( T) = \int d\omega \frac{\hbar \omega^3}{\pi^2c^2} \frac{1}{e^{\frac{\hbar\omega}{k_BT}}-1} \sigma_{\mathrm{abs}}(\omega)
\end{equation}
\noindent where $\hbar$, $k_B$ and $c$ are Planck's reduced constant, Boltzmann's constant and the speed of light, respectively. $P_{\mathrm{BB, abs}}$ is given by the same expression with fixed temperature $T_{0}=293$ K.
The absorption cross sections $\sigma_{\mathrm{abs}}$ are extrapolated from refractive index data provided for $\mathrm{Si}$  \cite{Franta2018} and $\mathrm{SiO_2}$  \cite{Franta2016}. Finally, the expression for $P_{\mathrm{gas}}$ is given by Ref. \cite{Chang2009}. Below the pressures considered here ($<10^{-8}$ mbar), this term in Eq. $\ref{eq:T_min}$ is negligible.

\begin{figure}[h]
    \centering
     \includegraphics{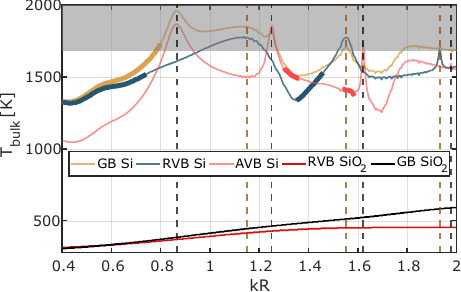}
    \caption{Bulk temperature of a spherical Si nanoparticle as a function of size parameter illuminated by GB, AVB and RVB. 
    Bold colors represent the stable trapping intervals for Si. The gray area indicates the Si melting temperature. }
\label{fig:Tnp_Si}
\end{figure}

Figure \ref{fig:Tnp_Si} shows the calculated temperature of the nanoparticle illuminated by all the beams discussed in this work, as a function of the size parameter, according to Eq. \ref{eq:T_min}. The thicker segments of the curves represent the regions where particles are trapped by the AVB, RVB and GB geometries, as discussed in sections V and VI. 

Higher temperatures were found for Si than Si$O_2$ in all trapping geometries. This is because: i) emission of black body radiation is much more efficient at temperatures below 1000 K for Si$O_2$ when compared to Si and ii)  the excitation of Mie resonances increases propagation times within the Si spheres which leads to larger absorption when compared to Rayleigh particles that have no resonances.  We also note that peaks in $\mathrm{Si}$ temperature are aligned with the Mie resonances excited by RVB (TM) and AVB (TE) modes, and that these local maxima in temperature can also be observed in the GB. Furthermore, both cylindrical polarized vortex beams perform better than GB for high refractive index nanoparticles (Si). Most importantly, the estimated temperatures are considerably lower than the melting temperature in the trapping interval (thicker lines). 

Temperatures seen in  Fig. \ref{fig:Tnp_Si} were calculated assuming $\mathrm{P}=\SI{500}{mW}$ leading to temperatures close to the $\mathrm{Si}$ melting point $\SI{1680}{K}$. 
Recent work \cite{Jnnemann2025} shows that the power absorption of Si at $\SI{1550}{\nano \meter}$ depends on internal temperature, $P_{\mathrm{abs}}(T)$, leading to heating above the limit we calculated even for lower power of $P=\SI{100}{mW}$. This is due to Si nonlinear absorption, which is mainly dominated by two photon, free carrier and bandgap absorption.
To mitigate these effects, we propose to use nanoparticles with similar refractive index but larger bandgap. AlGaAs is one such promising material which would allow trapping in high vacuum without melting.

\begin{figure}[h]
    \centering
    \includegraphics[]{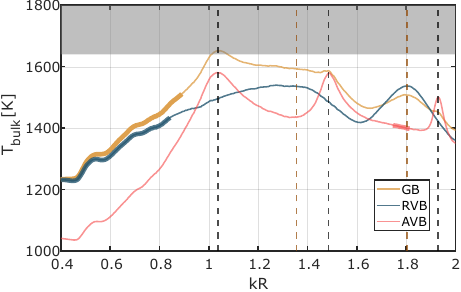}
    \caption{Bulk temperature of a spherical $\mathrm{Al_{0.9}Ga_{0.1}}As$ nanoparticle as a function of size parameter illuminated by GB, AVB and RVB. We assume a nonlinear absorption model (see Appendix C) and a laser power of $P=\SI{100}{mW}$. Bold colors represent the interval in which $\mathrm{Al_{0.9}Ga_{0.1}}As$ can be trapped in each type of beam. The gray area indicates $\mathrm{Al_{0.9}Ga_{0.1}}As$ melting temperature of $\SI{1640}{K}$.  }
    \label{fig:T_np_grouped_AlGaAs}
\end{figure}

Figure \ref{fig:T_np_grouped_AlGaAs} shows calculations of the bulk temperature of $\mathrm{Al_{0.9}Ga_{0.1}As}$ in high vacuum. The calculations take into account the temperature dependent two photon, free carrier and bandgap absorption for a laser power $P=\SI{100}{mW}$ as detailed in appendix C.  
The production of AlGaAs nanospheres have not been reported in the literature but other nanostructures, such as disks and cylinders, have been made for Mie-resonance applications \cite{Zhang2024, Vennberg2022}.

\section{Discussion and Conclusions}

We have shown that optical vector vortex beams created by counter propagating AVB, single AVB and RVB can be used to trap high refractive index particles (Si) in the Mie regime. Each is benchmarked against the conventional case of SiO$_2$ trapped in a GB, which remains the standard platform for optical levitation in high vacuum. Each of these geometries may be useful for different applications. The counter propagating AVB geometry provides the greatest suppression of recoil heating in the Mie regime by up to a factor of 10.6. However, this configuration is experimentally more demanding to implement. A simpler single AVB setup reduces recoil heating by a factor of 6.25 while also enabling the creation of both attractive and repulsive trapping potentials. The single beam RVB trapping is also relatively straightforward to implement experimentally with reduced recoil heating by up to a factor 9.2. However, it cannot be used to create repulsive potentials.   

Trapping high refractive index nanoparticles in both AVB and RVB provides a strong suppression of photon recoil heating when compared to the trapping of SiO$_2$ in GB. However, this comes at the cost of significant bulk heating of Si, which might preclude their use in some experiments. This trade-off motivates our proposal for an alternative material (AlGaAs) which has a refractive index closer to Si combined with a bandgap large enough to suppress the nonlinear absorption that will likely limit Si in levitation experiments.

AVBs can be useful for shaping the optical potential in the transverse direction. Here, by a suitable choice of wavelength/size parameter, we have shown that the potential can be tuned between harmonic and quartic and can even be made inverting. This versatility could also be useful for the range of protocols that are currently being explored to create non-classical states and deserves further study. 

Our analysis assumes that the particles are spherical. While generally nanoparticles used in levitation experiments are not completely spherical \cite{Rademacher2022}, nanoparticles of sufficient sphericity can be chosen after trapping by noting the absence of librational motion, as is currently done in experiments with colloidal silica nanoparticles \cite{Rademacher2022, Rademacher2023}. Further work will consider non-spherical particles in these traps.

The dark central region of the AVB may also be useful for trapping particles where doped atomic-like impurities are present. The utilization of these impurities is being considered for protocols that create non-classical states \cite{bose2017} or for the reduction of bulk heating by using laser refrigeration \cite{Rahman2017}.
Here the absence of an intense optical field at the center of the trapped particle, which is present in both RVB and GB, could be used to prevent undesirable optical pumping effects on the impurities.
\section{Disclosures}
The authors declare no conflict of interest.
\section{acknowledgments}
The authors acknowledge useful discussions with J.M.H. Gosling, M. Toro\v{s}, A. Pontin and F. Robicheaux and funding from the UK's EPSRC and STFC via Grant Nos. EP/N031105/1, EP/S000267/1, EP/W029626/1, EP/S021582/1 and ST/W006170/1.

\section*{Appendix A: Far field derivatives}

We use the Optical Tweezers Toolbox (OTT) to calculate the incident and scattered far field and the optical forces on a particle. As described in the main text, the toolbox allows a non-paraxial treatment of both the incident and scattered electromagnetic fields, making it particularly well-suited for the Mie scattering regime ($R \approx \lambda$).  We assume that the scattering fields propagate as $\bm{E_{scat}} = \bm{\mathcal{E}_{ k,k'}} (\theta, \phi) \frac{e^{ikr}}{r} $ and in the far field  where $\bm{E_{scat}} \cdot \bm{\hat{r}} = 0 $. As defined in the main text, $\bm{\mathcal{E}_{ k,k'}} (\theta, \phi) = {\mathcal{E}_{\theta}}  (\theta, \phi) \bm{\hat{\theta}} + {\mathcal{E}_{\phi}}  (\theta, \phi) \bm{\hat{\phi}}$ and we numerically calculate its derivative as
\begin{equation}
\begin{aligned}
        \frac{\partial {\mathcal{E}_{\theta}}  (\theta, \phi) }{\partial x_j} & \approx \frac{{\mathcal{E}_{\theta}}  (\theta, \phi; x_j+\delta x) - {\mathcal{E}_{\theta}}  (\theta, \phi; x_j-\delta x) }{2\delta x}  \\
 \frac{\partial {\mathcal{E}_{\phi}}  (\theta, \phi) }{\partial x_j} &\approx \frac{{\mathcal{E}_{\phi}}  (\theta, \phi; x_j+\delta x) - {\mathcal{E}_{\phi}}  (\theta, \phi; x_j-\delta x) }{2\delta x}
\end{aligned}
    \end{equation}
The derivatives used in Eq. \ref{eq:Mie_heating} of the main text are given by
\begin{subequations}
    \begin{align}
        \frac{\partial {\mathcal{E}}_x}{\partial x} = &\left ( \frac{\partial {\mathcal{E}_{\theta}}} {\partial x} {\hat{\theta}}  +  \frac{\partial {\mathcal{E}_{\phi}}   }{\partial x}  {\hat{\phi}} \right )  \cdot {\hat{x}} = \frac{\partial {\mathcal{E}_{\theta}}} {\partial x} \cos \theta\cos\phi - \frac{\partial {\mathcal{E}_{\phi}}   }{\partial x} \sin \phi \\
       \frac{\partial {\mathcal{E}}_y}{\partial x} = & \left ( \frac{\partial {\mathcal{E}_{\theta}}} {\partial x} {\hat{\theta}}  +  \frac{\partial {\mathcal{E}_{\phi}}   }{\partial x}  {\hat{\phi}} \right )  \cdot {\hat{y}} = \frac{\partial {\mathcal{E}_{\theta}}} {\partial x} \cos \theta\sin\phi + \frac{\partial {\mathcal{E}_{\phi}}   }{\partial x} \cos \phi \\
           \frac{\partial {\mathcal{E}}_z}{\partial x} = & \left ( \frac{\partial {\mathcal{E}_{\theta}}} {\partial x} {\hat{\theta}}  +  \frac{\partial {\mathcal{E}_{\phi}}   }{\partial x}  {\hat{\phi}} \right )  \cdot {\hat{z}} = - \frac{\partial {\mathcal{E}_{\theta}}} {\partial x} \sin \theta
    \end{align}
\end{subequations}

\section*{Appendix B: Direct comparison between radial vector and Gaussian beams}

\begin{figure}[h]
    \centering
    \includegraphics{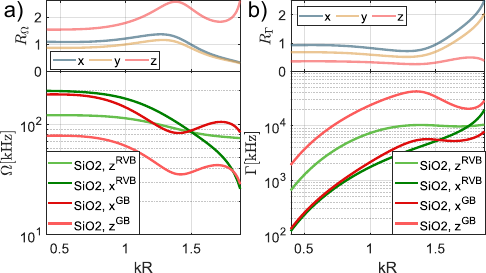}
    \caption{ Direct comparison of RVB and GB traps. a) Trap oscillation frequencies and b) recoil heating values (bottom) and ratio (top) for $\mathrm{SiO_2}$ in RVB traps with respect to GB trap. Oscillation frequency is higher and recoil heating is less for RVB in comparison to a GB, even for Rayleigh particles. The plots suggest the RVB trap could increase coherence time when compared to GB traps using $\mathrm{SiO_2}$ nanoparticles. } 
\label{fig:recoil_heating_radial_SiO2}
\end{figure}

Figure \ref{fig:recoil_heating_radial_SiO2}a shows the calculated oscillation frequencies for $\mathrm{SiO_2}$ nanospheres trapped in RVB and GB using the parameters of Table \ref{tab:parameters}. Also shown in the top panel are the ratios between these two trapping beams. Over the entire calculated interval, the oscillation frequency along the $z$ is higher for RVB traps than for GB traps, reaching a maximum enhancement factor of $2.56$ at $kR=1.38$. The improvement is a direct consequence of the tighter focusing of RVB relative to GB, as discussed in section \ref{sec:CP_beam}.

Figure \ref{fig:recoil_heating_radial_SiO2}b shows the calculated heating rates and ratios for both beams. The heating rates are lower for the RVB than for the GB in all three directions, until $kR  \approx 1.5$. After this limit, the transverse mechanical frequencies decrease considerably (see dark green in Figure \ref{fig:recoil_heating_radial_SiO2}a) 
as the particle size becomes comparable to the RVB focal spot. Consequently, the recoil heating in the x,y directions increases, as predicted by Eq. \ref{eq:recoil_heating} within the Rayleigh regime.

\section*{Appendix C: nonlinear absorption coefficient}

Figure \ref{fig:Tnp_Si} has been calculated using a first-order extinction coefficient $\alpha_0$, measured by absorption of light in Si at room temperature. 
However, as recently pointed out by Ref.\cite{Jnnemann2025}, nonlinear optical absorption at high bulk temperatures is dominant and has to be taken into account for Si. We follow Ref.\cite{Jnnemann2025} to define an effective absorption coefficient as: 
\begin{equation}
    \alpha_{eff} (T,I) [\mathrm{cm}^{-1}]= \alpha_0 + \alpha_{2PA} (T, I) + \alpha_{FCA}(T) + \alpha_{BG}(T) 
\end{equation}
\noindent where $\alpha_{2PA}$, $\alpha_{FCA}$, $\alpha_{BG}$ are two-photon absorption (2PA), free carrier absorption (FCA), and band-to-band (bandgap) absorption (BGA). These experimental numbers are well known for Si due to its high usage in integrated silicon-photonic devices in optical communications, and the resulting absorption coefficient is up to three orders of magnitude higher than $\alpha_0$, even at room temperature, for a typical trapping laser intensity \cite{Jnnemann2025}. 

For a more generic material, we follow Ref.\cite{VanStryland1986} to define 
\begin{equation}
    \alpha_{2PA}  = 3.1 \times 10^{2} \; \frac{ \sqrt{E_P} \frac{(\xi -1)^{3/2}}{\xi^5}} {n^2E_{gap}^3} \; I_0
    \label{eq:alpha_2PA}
\end{equation}
\noindent where $I_0$ is the laser intensity used here with units of $\mathrm{W / \mu m^2}$, $E_{p}$ is the Kane energy of a given material, $\xi = 2\hbar \omega/E_{gap}$ for $\hbar \omega=0.797\mathrm{ eV}$ being the photon energy and $\mathrm{E_{gap}}$ is the energy bandgap as a function of temperature defined as
\begin{equation}
    E_{\mathrm{gap}} (T) = E_{\mathrm{gap}, 0} - \frac{\alpha T^2}{T +\beta }
\end{equation}
\noindent with $E_{\mathrm{gap}, 0}$ being the energy bandgap at $\SI{0}{K}$, $\alpha$ and $\beta$ being empirical values for each material. 
The free carrier absorption coefficient is derived by classical Drude model as\cite{DRozario2022}
\begin{equation}
    \alpha_{FCA} = \frac{e^3 \lambda^2 N}{4\pi^2 \mu_H m_e^2 n \epsilon_0 c^3},
\end{equation}
\noindent where e is the electron charge, $N=\sqrt{N_C N_V} \exp [-E_{gap} (T)/2k_BT]$ is the carrier concentration for an effective density states in the conduction and valence band, $N_c$ and $N_v$, respectively, $\mu _H$ and $m$ being the mobility and the effective mass of the charge carrier. Bandgap absorption $\alpha_{BG}$ is modeled as discussed in Ref. \cite{Sturm1992}.
\section*{Appendix D: AlGaAs as an alternative}
\begin{figure}[h]
    \centering
    \includegraphics[]{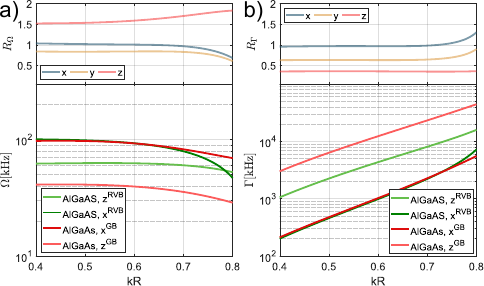}
    \caption{a) Oscillation frequencies and b) recoil heating
values (bottom) and ratio (top) for $\mathrm{Al_{0.9}Ga_{0.1}}As$ in RVB traps
with respect to GB trap. This plot has been shown here to illustrate the same features seen in Fig. \ref{fig:recoil_heating_radial_SiO2} for a less absorbing material.}
    \label{fig:recoil_heating_radial_AlGaAs}
\end{figure}

The necessity of increasing the bandgap energy to reduce nonlinear absorption while maintaining the high refractive index led us to $\mathrm{{Al_{x}Ga_{1-x}As}}$.  This III-V semiconductor is known to have a wide transparency window from $\SI{0.9}{\mu m}$ to $\SI{17}{\mu m}$ and an indirect bandgap absorption for $x>0.45 $ that allows us to model $\alpha_{BG}$ as in Ref.\cite{Sturm1992}. The fraction of aluminum (Al), x, creates different alloys whose main features are listed in Table \ref{tab:parameters_AlGaAs}. 
\begin{table}[h]
\centering
\caption{Table of parameters used in simulations }
\label{tab:parameters_AlGaAs}
\resizebox{\columnwidth}{!}{%
\begin{tabular}{l l}
\hline
\textbf{Parameters} & \textbf{Reference} \\
\hline
$\mathrm{n_{\mathrm{Al_{x}Ga_{1-x}As}} } = 3.26 - 0.42x - 0.027x^2 $ & Ref.\cite{Pavesi1994} \\
$\rho_{{\mathrm{Al_{x}Ga_{1-x}As}}} \; [ \mathrm{kg/m^3} ]= 5318 - 1560x $& Interpolation between AlAs and GaAs \\
$E_p \; [ \mathrm{eV} ] = 21+7x$& Interpolation between AlAs and GaAs \\
$E_{\mathrm{gap}, 0}\ ; [ \mathrm{eV} ] = 1.40x + 1.52$ & Fitting from Ref.\cite{Adachi1993-tj} \\
$\alpha \; [ \mathrm{eV/K} ] = (3.50x + 5.43) \times10^{-4} $ & Fitting from Ref.\cite{Adachi1993-tj} \\
$\beta \; [ \mathrm{K} ] = 124x + 209 $ & Fitting from Ref.\cite{Adachi1993-tj} \\
$N_c \; [ \mathrm{cm}^{-3} ]= 4.82\times10^{15} \; ((0.85-0.14x)T)^{3/2} $ & Ref.\cite{Ioffe} for $x>0.41$ \\
$N_v \; [ \mathrm{cm}^{-3} ]= 4.82\times10^{15} \;  ((0.51-0.25x)T)^{3/2} $ & Ref.\cite{Ioffe} for $x>0.41$ \\
\hline
\end{tabular}
}
\end{table}
A suitable choice of alloy is $\mathrm{{Al_{0.9}Ga_{0.1}As}}$. 
With an energy bandgap $E_{\mathrm{gap}} (300) = \SI{2.7}{eV}$ for $\mathrm{{Al_{0.9}Ga_{0.1}As}}$ being larger than that of Si ($\SI{1.1}{eV}$) at the same temperature, we approximate $\alpha_{BG}$ to be negligible over the temperature interval seen in Fig. \ref{fig:T_np_grouped_AlGaAs}.  
We follow the calculation routine described in the main text to present mechanical frequencies and recoil heating for $\mathrm{{Al_{0.9}Ga_{0.1}As}}$ at $P=\SI{100}{mW}$.
Figure \ref{fig:recoil_heating_radial_AlGaAs} presents an enhancement of the mechanical frequencies in the Rayleigh regime due to the use of RVB instead of GB for trapping. The higher mechanical frequencies decrease the recoil heating of RVB traps when compared to GB traps according to Eq. \ref{eq:recoil_heating}, following the same trend presented in Fig. \ref{fig:recoil_heating_radial_SiO2}. 
\begin{figure}[h]
    \centering
    \includegraphics[]{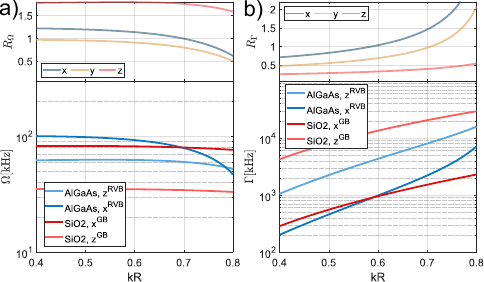}
    \caption{ a) Oscillation frequency and b) recoil heating values (bottom) and ratio (top) for $\mathrm{Al_{0.9}Ga_{0.1}}As$ in RVB traps with respect to the standard $\mathrm{SiO_2}$ in GB traps. Higher mechanical frequencies and less recoil heating can be obtained by using RVB in a high refractive index material than the standart $\mathrm{SiO_2}$ in GB traps.}
    \label{fig:recoil_heating_radial_AlGaAs_SiO2}
\end{figure}
A direct comparison with a low refractive index Si is seen in Fig. \ref{fig:recoil_heating_radial_AlGaAs_SiO2} where it is shown that an enhancement factor of 1.52 in the mechanical frequencies followed by a reduction of 2.85 in recoil heating can be reached when trapping $\mathrm{Al_{0.9}Ga_{0.1}}As$ in RVB beams instead of GB. 
\begin{figure}[h]
    \centering
    \includegraphics[]{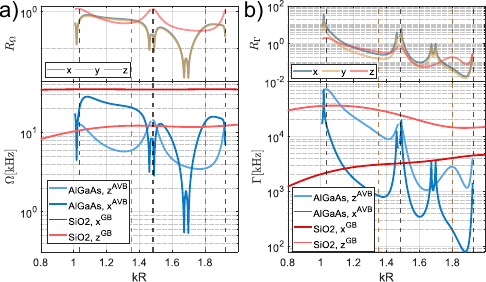}
    \caption{a) Oscillation frequency and b) recoil heating values (bottom) and ratio (top) for $\mathrm{Al_{0.9}Ga_{0.1}}As$ in counter propagating AVB and $\mathrm{SiO_2}$ in counter propagating GB, both at low numerical aperture (NA=0.4).}
    \label{fig:recoil_heating_AVB_cp_AlGaAs}
\end{figure}
Lastly, Fig. \ref{fig:recoil_heating_AVB_cp_AlGaAs} shows the oscillation frequencies and recoil heating for counter propagating AVB at $\mathrm{NA}=0.4$. A small interval for single AVB trapping is also found for high NA=0.8 as seen in Fig. \ref{fig:T_np_grouped_AlGaAs}.
\newpage
\bibliography{main.bib}
\end{document}